\documentstyle[12pt]{article}
\begin{document}

\begin{center}
{\LARGE{\bf Differential calculus on $q$-Minkowski space\footnote{
To appear in the Proc. of the $XXX$-th Karpacz Winter School of
Theoretical Physics (Feb. 1994) (Polish Sci. Pub., J. Lukierski,
Z. Popovicz and J. Sobczyk eds.).}}}
\end{center}

\begin{center}
{\large{J.A. de Azc\'{a}rraga$^{*}$ and F. Rodenas$^{* \dag}$}}
\end{center}
\addtocounter{footnote}{-1}

\begin{center}
{\small{\it $* \quad$ Departamento de F\'{\i}sica Te\'{o}rica and IFIC,\\
Centro Mixto Universidad de Valencia-CSIC\\
E-46100-Burjassot (Valencia), Spain.}}
\end{center}

\begin{center}
{\small{\it $ \dag \quad$ Departamento de Matem\'atica Aplicada,\\
Universidad Polit\'ecnica de Valencia \\
E-46071 Valencia, Spain.}}
\end{center}

\vspace{1\baselineskip}

\indent
We wish to report here on a recent approach \cite{AKR} to the non-commutative
calculus
on $q$-Minkowski space which is based on the reflection equations (RE) with
no spectral parameter.
These are considered as the expression of the invariance (under the
coaction of the $q$-Lorentz group) of the commutation properties which define
the different $q$-Minkowski algebras. This approach also allows us to
discuss the
 possible ambiguities in the definition of
$q$-Minkowski space ${\cal M}_q$ and its differential
calculus. The
commutation relations  among  the generators of
${\cal M}_q$ (coordinates), ${\cal D}_q$ (derivatives), $\Lambda_q$
(one-forms) and a few invariant
(scalar) operators  are established.
The correspondence of the obtained expressions with those of \cite{OSWZ-CMP}
is  given in a table at the end.

\section{Covariance, the $q$-Lorentz group and the RE}

\indent
As  is well known, the classical 2$\rightarrow$1 homomorphism
$SL(2,C)/Z_2 \simeq L^{\uparrow}_+$ is based on the
observation that the transformation of the hermitian matrix $K$,

\begin{equation}\label{dosb}
K \longmapsto A K \tilde{A}^{-1} \equiv  A K A^{\dagger} =K'\quad ,
\end{equation}

\noindent
necessarily produces another  hermitian
matrix $K'$ and that $detK=detK'$ if $detA=1$.
If
$\sigma^{\mu}=
(\sigma^{0}, \sigma^{i})$ (where $\sigma^{0}$=${\bf I}$
and $\sigma^{i}$ are the Pauli
matrices), $K$ may be written as $K = \sigma^{\mu} x_{\mu}$ and
$det K = (x^{0})^{2} - \vec{x}^{2}$.
Thus, $K' =
\sigma^{\mu} x_{\mu}'$ with $x'^{\mu} = \Lambda^{\mu} _{\,\,\nu} x^{\nu}$,
and the
correspondence $\pm A \mapsto \Lambda$$\in$$L_+^{\uparrow}$
realizes the  covering
homomorphism of the restricted Lorentz group. The antisymmetric matrix
$\epsilon \equiv
i \sigma^{2}$ realizes the equivalence of  the
representations $A$ and $(A^{-1})^t$ of $SL(2,C)$,
$\epsilon A \epsilon^{-1} = (A^{-1})^{t}$;
hence
\begin{equation}\label{dosc}
\epsilon A^{*} \epsilon^{-1} = (A^{-1})^{\dagger} \equiv \tilde{A}\quad,\quad
K'^{\epsilon} = \tilde{A} K^{\epsilon} A^{-1} \quad , \quad K^{\epsilon}
\equiv \epsilon K^{*} \epsilon^{-1} \;.
\end{equation}
\noindent
Clearly, since $\epsilon(\sigma^{\mu})^*\epsilon^{-1}=(\sigma^0,-\sigma^i)
\equiv \rho^{\mu}$, another set of hermitian matrices
(not related by a similarity transformation to $\sigma^{\mu}$) may be
introduced.
All this reflects that for $SL(2,C)$ the representations $A$ and $\tilde{A}
=\epsilon A^* \epsilon^{-1}$ are unequivalent since the complex conjugation
$*$ is an external automorphism; $A$ and $A^*$ define the two fundamental
representations under which undotted and dotted spinors transform.
In contrast, the vector representation $D^{\frac{1}{2},\frac{1}{2}}$
is real and the $*$-conjugation, which in general produces unequivalent
representations ($*: D^{j,j'} \mapsto D^{j',j}$), acts trivially
on it ($K$ is hermitian).

The crucial idea \cite{PW,WA-ZPC48,WA-A6,SWZ-ZPC52,OSWZ-CMP,SONG}
to deform the Lorentz group was to replace the
$SL(2,C)$ matrices $A$  by the generator matrix $M$ of the quantum
group $SL_{q}(2,C)$. Due to the fact that the hermitian conjugation
$(M^{\dagger} _{ij} = M_{ji} ^{*})$ includes the $*$-operation, an extra copy
$\tilde{M}$ of $SL_{q}(2,C)$ was introduced, with entries not commuting with
those of $M$. The $R$-matrix form of the commutation relations among the
quantum group generators $(a, b, c, d)$ of $M$ and $(\tilde{a}, \tilde{b},
\tilde{c}, \tilde{d})$ of $\tilde{M}$ may be  expressed by
\begin{equation}\label{dosf}
\begin{array}{l}
R_{12} M_{1} M_{2} = M_{2} M_{1} R_{12}\quad,\\
 R_{12} \tilde{M}_{1} \tilde{M}_{2} = \tilde{M}_{2} \tilde{M}_{1}
R_{12}\quad,\\
R_{12} M_{1} \tilde{M}_{2} = \tilde{M}_{2} M_{1} R_{12}\quad.
\end{array}
\end{equation}
\noindent
As in the classical
 case (eq. (\ref{dosc})),
here $M$ and $\tilde{M}$ are related
by requiring $\tilde{M}^{-1}
=M^{\dagger}$.
This condition is consistent with all relations
in (\ref{dosf}) provided that the
deformation parameter $q$ is real, $q \in {\bf R}$. The
set of  generators $(a, b, c,
d, \tilde{a}, \tilde{b}, \tilde{c}, \tilde{d})$ satisfying
$det_{q} M$=1=$det_{q} \tilde{M}$, the commutation relations (\ref{dosf}) and
the conditions
$\tilde{M}^{-1}
=M^{\dagger}$  define the quantum Lorentz group $ L_{q}$.
There are, however,  other  possibilities
if we allow for different
$R$ matrices in  (\ref{dosf}) reflecting different commutation properties.

To introduce the
$q$-Minkowski {\it algebra} ${\cal M}_{q}$ it is natural to extend
(\ref{dosb})
to the quantum case by stating that the algebra generated by the entries of
$K$ is a comodule for the coaction $\phi$ defined by
\begin{equation}\label{dosh}
\phi : K \longmapsto
K' = M K \tilde{M}^{-1} \,, \quad K'_{is} = M_{ij} \tilde{M}^{-1}_{ls}
K_{jl} \equiv  \Lambda_{il,js} K_{jl} \,,
\end{equation}

\noindent
where it is assumed that the matrix elements of $K$ commute with those of $M$
and $\tilde{M}$ but not among themselves.
Much in the same way as the commuting properties of $q$-two-vectors
(or, better said here, $q$-spinors) are preserved by the coaction of $M$
and $\tilde{M}$, we now demand that the commuting properties of the entries
of $K$  are preserved by
(\ref{dosh}). More specifically, in order to identify the elements of $K$
\begin{equation}\label{dosi}
K= \left(
\begin{array}{ll}
\alpha & \beta\\
\gamma & \delta
\end{array}
\right)
\end{equation}

\noindent
with the generators of  ${\cal M}_{q}$ we require,
as in the classical case,

\noindent
a) \, a reality property preserved by (\ref{dosh}),

\noindent
b) \,  a set of commutation relations for  the elements of $K$
(a `presentation' of the

algebra ${\cal M}_q$) preserved by
(\ref{dosh}),

\noindent
c) \,  a (real) $q$-Minkowski length $l_q$, defined through the
$q$-determinant $det_qK$ of  $K$,
invariant under the $q$-Lorentz transformation (\ref{dosh}).

The reality condition  $K= K^{\dagger}$ (a) is consistent with (\ref{dosh})
since $\tilde{M}^{-1} = M^{\dagger}$ as in the classical case.
Let us now explore  (b)  to express the six basic relations for
the elements of $K$ which define a $q$-Minkowski algebra.

 We may describe them
\cite{AKR} by means of a
general RE  (see \cite{K-SKL,KS} and references therein; see also
\cite{MAJ-LNM,SM2}
in the context of braided algebras)
\begin{equation}\label{dosak}
R^{(1)} K_{1} R^{(2)} K_{2} = K_{2} R^{(3)} K_{1} R^{(4)}.
\end{equation}

\noindent
The four $4 \times 4$ $ R^{(i)}$ matrices have now to be found  by demanding
the invariance of (\ref{dosak}) under (\ref{dosh}). From
 $R^{(1)} K'_{1} R^{(2)} K'_{2} = K'_{2} R^{(3)} K'_{1} R^{(4)}$
and (\ref{dosh}) we obtain
\begin{equation}
R^{(1)} M_1K_{1}\tilde{M}_1^{-1} R^{(2)}M_2 K_{2}\tilde{M}_2^{-1}
= M_2K_{2} \tilde{M}_2^{-1}R^{(3)} M_1K_{1} \tilde{M}_1^{-1}R^{(4)}
\end{equation}
\noindent
an expression which may be written as
\begin{equation}
R^{(1)} M_1M_2 K_{1} R^{(2)} K_{2} \tilde{M}_1^{-1}\tilde{M}_2^{-1}
= M_2M_1K_{2} R^{(3)} K_{1} \tilde{M}_2^{-1} \tilde{M}_1^{-1}R^{(4)}
\end{equation}
\noindent
if
\begin{equation}\label{dosal1}
R^{(2)} M_{2}
\tilde{M}_{1} = \tilde{M}_{1} M_{2} R^{(2)}\quad,\quad
R^{(3)} M_{1} \tilde{M}_{2} = \tilde{M}_{2} M_{1} R^{(3)} \quad ,
\end{equation}
\noindent
and which is equal again to (\ref{dosak}) if
\begin{equation}\label{dosal}
R^{(1)} M_{1} M_{2} = M_{2} M_{1} R^{(1)} \quad , \quad
R^{(4)} \tilde{M}_{2} \tilde{M}_{1} = \tilde{M}_{1} \tilde{M}_{2} R^{(4)}\;.
\end{equation}

\noindent
Comparing these
equations (\ref{dosal1}) and (\ref{dosal}) with (\ref{dosf})
we find that the $q$-Lorentz group (\ref{dosf})
leads to
\begin{equation}\label{dosan}
R^{(1)} = R_{12} \,\mbox{or}\, R_{21}^{-1} \;, \quad R^{(2)} = R_{21}\;, \quad
R^{(3)} = R_{12}\;, \quad R^{(4)} = R_{21} \,\mbox{or} \, R_{12}^{-1}\,.
\end{equation}

The solution  $R^{(1)} $=$ R_{12}$, $ R^{(4)} $=$ R_{21}$
gives
\begin{equation}\label{re}
R_{12}K_1R_{21}K_2= K_2R_{12}K_1R_{21} \quad.
\end{equation}

\noindent
For $K$ expressed as in  (\ref{dosi}), eq. (\ref{re}) is equivalent to
the six basic relations \cite{WA-ZPC48,SWZ-ZPC52,MAJ-LNM}
\begin{equation}\label{dosl}
\begin{array}{lll}
\alpha \beta = q^{-2} \beta \alpha \quad ,\quad &
\alpha \gamma = q^{2} \gamma \alpha \quad,  \quad  & {} [\alpha , \delta ]
= 0 \quad ,\\
{} [\beta , \gamma] = q^{-1} \lambda (\delta - \alpha ) \alpha \quad ,\quad &
{} [\delta, \beta ] = q^{-1} \lambda \alpha \beta \quad ,\quad &
{} [\gamma , \delta ] = q^{-1} \lambda \gamma \alpha \quad ,
\end{array}
\end{equation}

\noindent
which characterize ${\cal M}_{q}$.
We shall  adopt the point of view that this associative,
non-commutative algebra or `quantum space'
is the primary object on which the non-commutative differential calculus
will be constructed.
We may then give the following

\noindent
{\it Definition}

The quantum Minkowski space-time algebra is the
non-commutative associative algebra
${\cal M}_{q}$ generated by the four elements of $K$, subjected to the
reality conditions $\alpha = \alpha^{*} \,,\, \delta = \delta^{*}
\,,\, \beta^{*} = \gamma \,,\, \gamma^{*} = \beta$, and satisfying the
commutation relations  (\ref{dosl}) defined by (\ref{re}).

The central (commuting) elements of ${\cal M}_{q}$ may be obtained by using
the $q$-trace $tr_{q}$ \cite{FRT1,ZU-MPX}. The linear one is the $q$-trace
of $K$
\begin{equation}\label{dosr}
c_{1} \equiv tr_{q} K \equiv  tr(DK)= q^{-1} \alpha + q \delta \quad, \quad
D= diag(q^{-1}, q)= \epsilon^q \epsilon^{qt} \;
\end{equation}
\noindent
($\epsilon^q$ is the $q$-antisymmetric tensor)
and satisfies $tr_q{\cal O}=tr_q(M{\cal O} M^{-1})$
since
$M^tD(M^{-1})^t$ \\ $=D$. The higher order central elements are given by
\begin{equation}\label{dosv}
c_{n} \equiv tr_{q} K^{n} \quad , \quad K c_{n} = c_{n} K    \quad .
\end{equation}
\noindent
The $q$-trace $tr_{q} K = c_{1}$ is central but not invariant.
As in the classical case, where $1/2\,tr(\sigma_{\mu}x^{\mu})= x^0$,
we may identify $1/[2]\,tr_qK$ with the time coordinate.
The first two central elements  $c_{1}$ and $c_{2}$ are algebraically
independent;
 for $n > 2$ the $c_{n}$ are polynomial functions of them due to the
characteristic equation for $K$\, \cite{K-SKL,KS},
\begin{equation}\label{dosw}
q K^{2} - c_{1} K + \frac{q}{[2]} (q^{-1} c^{2}_{1} - c_{2}) I= 0\,.
\end{equation}

The $q$-determinant (condition c) $det_{q} K$ of $K$ is obtained by means of
the
$q$-antisym\-metri\-zer $P_{-}$, which is a rank one $\,4 \times 4$
projector. It is defined \cite{K-SKL} through
\begin{equation}\label{dosx}
(det_{q} K) P_{-} = -q P_{-} K_{1} \hat{R} K_{1} P_{-} = (\alpha \delta -
q^{2} \gamma \beta ) P_{-}\,,
\end{equation}

\noindent
where $\hat{R}={\cal P}R$ with ${\cal P}$ the permutation operator.
We now  identify this real, central (since
$det_qK=q^2(q^{-1}c_1^2-c_2)/[2]$) and
invariant element with the square $l_q$ of the
$q$-Minkowski invariant length \cite{WA-ZPC48,SWZ-ZPC52,SONG,OSWZ-CMP}
\begin{equation}\label{dosaa}
l_{q} \equiv det_qK = \alpha \delta - q^{2} \gamma \beta \;,
\qquad l_{q} \in {\cal M}_{q}\,.
\end{equation}

 If $K$ transforms
by (\ref{dosh}) say, contravariantly, then
\begin{equation}\label{dosaf}
K_{ij}^{\epsilon} = \hat{R}^{\epsilon}_{ij,kl} K_{kl} \quad \quad
(K^{\epsilon} = \hat{R}^{\epsilon} K) \quad,
\end{equation}

\noindent
where $ \hat{R}^{\epsilon} = ({\bf 1} \otimes \epsilon^{qt})
\hat{R} ({\bf 1} \otimes (\epsilon^{q\;-1})^{t})$,
transforms covariantly  {\it i.e.} $K^{\epsilon '}
= \tilde{M} K^{\epsilon} M^{-1}$
(with our definition of $\hat{R}^{\epsilon}$, $K^{\epsilon}$ reduces for
$q$=1 to $K^{\epsilon}$ in (\ref{dosc}) but for a sign).
When
$det_qK \neq 0$,  $K^{\epsilon} $=$ -q^{-1} (det_{q} K) K^{-1} $.
The `length' of a  $q$-Minkowski vector
is then  given by the central $L_q$-invariant $q$-trace
of $K^{\epsilon}K$
\begin{equation}\label{dosah}
l_{q} \equiv  det_{q} K = \frac{-q}{[2]} tr_{q} (K K^{\epsilon})=
\frac{-q}{[2]}
tr_{q} (K^{\epsilon} K) \quad ,\quad [l_q,K]=0\;.
\end{equation}

\noindent
Let us now introduce a $q$-Minkowski tensor $g_{ij,kl}$ by means of the
expression
\begin{equation}\label{metric}
g_{ij,kl}
=\frac{-q^{-1}}{\left[2\right]} D_{si} \hat{R}_{js,kl}^{\epsilon}
= \frac{q^{-1}}{\left[2\right]} \epsilon^q_{im}
\hat{R}_{jm,kt} \epsilon^{q\;-1}_{lt}\quad, \quad
(g=\frac{-q^{-1}}{\left[2\right]}D_1 {\cal P} \hat{R}_{12}^{ \epsilon}) \quad,
\end{equation}
\noindent
so that
\begin{equation}\label{ch}
l_q=\frac{-q}{\left[2\right]} D_{si} \hat{R}_{js,kl}^{\epsilon}
K_{ij} K_{kl}=q^{2} g_{ij,kl} K_{ij} K_{kl}
\quad.
\end{equation}
\noindent
Since
\begin{equation}\label{metric2}
\Lambda^t g \Lambda = g \quad, \quad
\Lambda_{rs,ij} g_{rs,mn} \Lambda_{mn,kl}=g_{ij,kl}\;,
\end{equation}
\noindent
the tensor $g_{ij,kl}$ may be identified with the $q$-Minkowski metric.

\vspace{1\baselineskip}
Let us come back to the other solutions of (\ref{dosak}) in (\ref{dosan}).
The possibility $R^{(1)} $=$ R_{12}$,  $ R^{(4)} $=$ R_{12}^{-1}$ implies
replacing (\ref{re}) by
\begin{equation}\label{dosap}
R_{12} K_{1} R_{21} K_{2} = q^{2} K_{2} R_{12} K_{1} R_{12}^{-1}\,,
\end{equation}

\noindent
which defines a {\it new} $K$.
The commutation properties of its
entries, however,  are again given by
(\ref{dosl}) although  now
$det_qK=0$. Indeed, multiplying (\ref{dosap}) by $P_-{\cal P}$ from the left
and by ${\cal P}P_-$ from the right, we obtain using
$\hat{R}^{\pm 1}= q^{\pm 1}P_+ - q^{\mp 1}P_-$ that
 $(1-q^4)P_-K_1 \hat{R}K_1P_-=0$  and ($q^4 \neq 1$) $det_qK=0$. Also, the
insertion of the $R$-matrix identity $q^2 R_{12}^{-1}$=$ R_{21}-
q \lambda [2]P_-{\cal  P}$ in eq. (\ref{dosap}) reproduces (\ref{re}),
\begin{equation}\label{at2}
R_{12} K_{1} R_{21} K_{2} =  K_{2} R_{12} K_{1} R_{21}
-q\lambda [2] K_2 R_{12}K_1 P_- {\cal P} =K_{2} R_{12} K_{1} R_{21}\,,
\end{equation}
\noindent
since $det_qK=0$. This means that nothing is gained by
considering (\ref{dosap}) as a separate case, and
it may be discarded.
The other two solutions $R^{(1)} $=$ R_{21}^{-1}$, $ R^{(4)} $=$
R_{12}^{-1}$ and $R^{(1)} $=$ R_{21}^{-1}$, $ R^{(4)} $=$ R_{21}$
may be easily seen, respectively,
the same as (\ref{re}) and (\ref{dosap}); thus, if the $q$-Lorentz
group is defined by eqs. (\ref{dosf}) we are led  uniquely to
(\ref{re}) or (\ref{dosl}) as
the relations defining the $q$-Minkowski algebra ${\cal M}_{q}$.

\section{\bf Deformed derivatives and $q$-De Rham complex}

\indent
The development of a non-commutative differential calculus
(see, {\it e.g.}, \cite{SLWO,JWBZ,ZU-MPX,MANIN} and references therein)
requires including derivatives and differentials. We shall now do this
\cite{AKR}
for the $q$-Minkowski space  by extending
the RE to accommodate them appropriately;
in this way, the $q$-derivatives (${\cal D}_q$) and
the $q$-forms ($\Lambda_q$)  algebras will be defined by RE.
Consider first an object $Y$ transforming covariantly {\it i.e.},
\begin{equation}\label{ta}
Y \longmapsto Y' = \tilde{M} Y M^{-1} \quad ,
\quad Y =
\left[
\begin{array}{ll}
u & v\\
w & z
\end{array}
\right]\;.
\end{equation}
\noindent
Proceeding now as for (\ref{dosak}) with $Y$ replacing $K$,
the invariance of the commutation properties of the matrix elements of $Y$
 gives
\begin{equation}\label{tb}
R^{(1)} = R_{12}\, \mbox{or}\, R_{21}^{-1} \,, \quad R^{(2)}=R_{12}^{-1} \,,
\quad R^{(3)}=R_{21}^{-1} \,,\quad  R^{(4)}= R_{21} \,\mbox{or}\, R_{12}^{-1}
\,.
\end{equation}

\noindent
These four possibilities again reduce to two,
\begin{eqnarray}
R_{12} Y_{1} R_{12}^{-1} Y_{2} = Y_{2} R_{21}^{-1} Y_{1}
R_{21} \quad, \label{tc} \\
q^{2} R_{21}^{-1} Y_{1} R_{12}^{-1} Y_{2}
                         = Y_{2} R_{21}^{-1} Y_{1} R_{21}\quad,\label{td}
\end{eqnarray}
\noindent
of which we shall retain only (\ref{tc}) since (\ref{td}) leads
(in similarity with (\ref{dosap})) to
the same algebra plus the condition $det_qY$=0 (see (\ref{tf}) below).

The (central and $q$-Lorentz invariant) $q$-determinant is defined through
\begin{equation}\label{tf}
(det_{q} Y) P_{-} =(-q^{-1}) P_{-} Y_{1} \hat{R}^{-1} Y_{1} P_{-}
= (uz - q^{-2} v w) P_-\quad.
\end{equation}

\noindent
Since $Y$ is covariant, we may define (cf. (\ref{dosaf})) a contravariant
$Y^{\epsilon}$ by
$Y^{\epsilon}= (\hat{R}^{\epsilon})^{-1} \, Y $
(when $det_{q} Y \neq 0 , \; Y^{\epsilon} = -q (det_{q} Y) Y^{-1})$; then,
(cf. (\ref{dosah}))
\begin{equation}\label{ti}
\Box_q \equiv  det_{q} Y = - \frac{q^{-1}}{[2]} tr_{q} (Y Y^{\epsilon}) = -
\frac{q^{-1}}{[2]}
tr_{q} (Y^{\epsilon} Y) \;, \quad [ \Box_q , Y]=0\;,
\end{equation}
\noindent
where ${\Box}_q$ becomes the central and $L_q$-invariant  $q$-D'Alembertian
once the components of $Y$ are associated with the $q$-derivatives.
As the $K$ matrix entries were associated with the generators of
${\cal M}_q$, we shall consider the elements of $Y$ as those of
the algebra ${\cal D}_q$
of the $q$-Minkowski derivatives.

The next step in constructing the non-commutative
$q$-Minkowski differential calculus is to establish the commutation
properties among coordinates
and derivatives. We need  extending  the classical
relation $\partial_{\mu} x^{\nu} = x^{\nu} \partial_{\mu} + \delta_{\mu}^{\nu}
$, $\partial^{\dagger}$=$- \partial$,  to the non-commutative
case in a $q$-Lorentz invariant way. This may be done by means of
 an {\it inhomogeneous} RE of the form \cite{AKR}
\begin{equation}\label{tk}
Y_{2} R^{(1)} K_{1} R^{(2)} = R^{(3)} K_{1} R^{(4)} Y_{2} + \eta J,
\end{equation}

\noindent
where $\eta$ is a constant, $\eta J$$\rightarrow$$I_{4}$ in the
$q$$ \rightarrow$$1$ limit, and $J$ is invariant,
\begin{equation}\label{tl}
J \longmapsto \tilde{M}_{2} M_{1} J \tilde{M}_{1}^{-1} M_{2}^{-1} = J
\,, \qquad  \tilde{M}_{2} M_{1} J = J M_{2} \tilde{M}_{1}\,.
\end{equation}

\noindent
This equation exhibits the need of having $K$ and $Y$ transforming {\it e.g.}
contravariantly and covariantly;
indeed, the assumption that they transform in the same manner (both as $K$,
say)
leads to $M_{1} M_{2} J = J \tilde{M}_{1} \tilde{M}_{2}$ which cannot be
fulfilled  already in the $q=1$ case since it would imply the equivalence of
unequivalent representations.

Again, an analysis similar to those of Sec.1 leads to
\begin{equation}\label{tm}
R^{(1)} = R_{12}\,\mbox{or}\,  R_{21}^{-1} \,, \quad R^{(2)} = R_{21} \,,
\quad  R^{(3)} =
R_{12} \,,\quad  R^{(4)} = R^{-1}_{12} \,\mbox{or}\,  R_{21} \, .
\end{equation}
\noindent
As for $J$,  setting $J\equiv  J' {\cal P}$ in eq. (\ref{tl}) gives
$\tilde{M}_{2}
M_{1} J' = J' M_{1} \tilde{M}_{2}$, hence $J = R_{12} {\cal P}$ (the same
result follows if  we set $J= {\cal P} J'$). This means that there are, in
principle, four basic possibilities consistent with covariance expressing the
commutation properties of coordinates (elements of $K$) and derivatives
(entries of $Y$). These read
\begin{eqnarray}
Y_{2} R_{12} K_{1} R_{21} = R_{12} K_{1} R^{-1}_{12} Y_{2} + \eta_{1}
R_{12}{\cal P}\quad;\label{tn} \\
Y_{2} R_{21}^{-1} K_{1} R_{21} = R_{12} K_{1} R_{21} Y_{2} + \eta_{2}
R_{12}{\cal P}\quad; \label{to} \\
Y_{2} R_{12} K_{1} R_{21} = R_{12} K_{1} R_{21} Y_{2} + \eta_{3}
R_{12}{\cal P}\quad; \label{tp} \\
Y_{2} R_{21}^{-1} K_{1} R_{21} = R_{12} K_{1} R^{-1}_{12} Y_{2} + \eta_{4}
R_{12}{\cal P}\quad. \label{tq}
\end{eqnarray}
\noindent
Due to the fact that these expressions now involve $K$ and $Y$, they are all
unequivalent. In fact, we do not need assuming that the four $Y's$ appearing
in each of the equations (\ref{tn}-\ref{tq}) are the same; all
that it is demanded is that  they all transform as in (\ref{ta}).

Let us now look at the hermiticity properties of $K$ and $Y$. It is clear that,
since $R^t = R^{\dagger}$ ($q$ is real), eqs. (\ref{re}) and (\ref{tc}) are
independently consistent
with the hermiticity of  $K$ and the antihermiticity of $Y$. However, this is
no longer the case
if the  inhomogeneous equations are included. Keeping the physically reasonable
assumption that $K$ is hermitian, eq. (\ref{tn}) gives
\begin{equation}\label{tr}
Y_{2}^{\dagger} R_{21}^{-1} K_{1} R_{21} = R_{12} K_{1} R_{21} Y_{2}^{\dagger}
 - \eta_{1} R_{12}{\cal P}\quad;
\end{equation}

\noindent
{\it i.e.}, $Y^{\dagger}$ satisfies the commutation relations  given by
the second inhomogeneous equation (\ref{to}) for $\eta_{2} $$=$$ - \eta_{1}$
(of
course, $Y^{'\dagger} = \tilde{M} Y^{\dagger} M^{-1}$ again since $\tilde{M} =
(M^{-1})^{\dagger})$. Thus, we need accommodating $Y^{\dagger}$ by means of
{\it another} reflection equation, eq. (\ref{to}) for $Y^{\dagger}$.
Having
selected (\ref{tn}) for $Y$  and (\ref{to}) for $Y^{\dagger}$,
we may now consider the  possibilities
(\ref{tp}) or  (\ref{tq}). It turns out that they are inconsistent with
the previous relation (\ref{re}),
what may be seen  with a little effort by acting on (\ref{re})
 with an additional  $Y$ and using (\ref{tp}) or
(\ref{tq}). For instance, multiplying (\ref{re}) from the left by
$Y_3R_{23}R_{13}$ and using twice (\ref{tp}) and the YBE for different ordering
of the $R$-matrices, we obtain, respectively, for the left and right hand sides
\begin{equation}\label{a}
\mbox{l.h.s.}: \begin{array}{l}
R_{23}R_{13}K_2R_{12}K_1R_{21}R_{31}R_{32}Y_3R_{32}^{-1}R_{31}^{-1} \\
+ \eta_3 R_{12}R_{13}K_1R_{31}{\cal P}_{32}
+ \eta_3 R_{12}{\cal P}_{13}R_{23}R_{21}K_2\quad,
\end{array}
\end{equation}

\begin{equation}\label{b}
\mbox{r.h.s.}: \begin{array}{l}
R_{23}R_{13}K_2R_{12}K_1R_{21}R_{31}R_{32}Y_3R_{32}^{-1}R_{31}^{-1} \\
+ \eta_3 {\cal P}_{23}R_{13}R_{12}K_1 R_{21}
+ \eta_3 R_{23}K_2R_{32}{\cal P}_{13}R_{21}\quad.
\end{array}
\end{equation}

\noindent
The cubic  and the first linear terms coincide, and   the last ones can
be rewritten, respectively, as
\begin{equation}\label{c}
\eta_3 R_{12}R_{21}({\cal P}_{13}R_{21}K_2) \quad , \quad
\eta_3 ({\cal P}_{13}R_{21}K_2) R_{12}R_{21} \quad .
\end{equation}

\noindent
Now, since   $R_{12}R_{21}=I + \lambda R_{12}{\cal P}$,
we conclude that the two  terms in  (\ref{c}) are different.
Then, the relation (\ref{tp})  (and, analogously, (\ref{tq}))
is inconsistent with (\ref{re}).
In contrast, similar calculations show that (\ref{tn}) and (\ref{to}) are
consistent
with the commutation properties of ${\cal M}_q$ and ${\cal D}_q$.

In order to have the inhomogeneous term in the simplest form (the analogue of
the $\delta_{\nu}^{\mu}$ of the $q=1$ case) it is convenient to take $\eta_{1}
= q^{2}$ and to redefine $Y^{\dagger}$ as $\tilde{Y} = -q^{-4}
Y^{\dagger}$. In this way, the equations describing the
commutation relations of the generators of the algebras of coordinates $(K)$,
derivatives $(Y)$ and their hermitian conjugates $(Y^{\dagger}
\propto \tilde{Y})$ are given by
\begin{equation}\label{ts}
\begin{array}{l}
Y_{2} R_{12} K_{1} R_{21} = R_{12} K_{1} R^{-1}_{12} Y_{2} + q^2
R_{12}{\cal P} \quad, \\
\tilde{Y}_{2} R_{21}^{-1} K_{1} R_{21} = R_{12} K_{1} R_{21} \tilde{Y}_{2}
+ q^{-2} R_{12}{\cal P}\quad.
\end{array}
\end{equation}

\noindent
Notice that, although we  identified $Y$ with the derivatives and $\tilde{Y}$
with their hermitians, the reciprocal assignment is also possible.
We may also introduce a RE for $Y$ and $\tilde{Y}$;
consistency selects from (\ref{tb}) the solution
\begin{equation}\label{yy}
\tilde{Y}_2R_{21}^{-1} Y_1 R_{21}= R_{21}^{-1} Y_1 R_{12}^{-1}\tilde{Y}_2
\quad.
\end{equation}

The determination of the commutation relations for the $q$-De Rham complex
requires
introducing the exterior derivative $d$  \cite{OSWZ-CMP};
we shall assume that $d^{2}$=$0$
and that it satisfies the Leibniz rule. To the
four generators
of the ${\cal M}_{q}$ (entries of $K$) and of ${\cal D}_{q}$ ($Y$) Minkowski
algebras we now add the four elements  of $dK$
($q$-one-forms), which generate the
De Rham complex algebra  $\Lambda_{q}$
(the degree of a form is defined as in the classical case). Clearly,  $d$
commutes with the
$q$-Lorentz coaction (\ref{dosh}), so that
\begin{equation}\label{tt}
dK' = M d K \tilde{M}^{-1} \quad .
\end{equation}

\noindent
Applying $d$ to  (\ref{re}) we obtain
\begin{equation}\label{tu}
R_{12} dK_{1} R_{21} K_{2} + R_{12} K_{1} R_{21} d K_{2} = d K_{2} R_{12}
K_{1}R_{21} + K_{2} R_{12} d K_{1} R_{21} \quad .
\end{equation}

\noindent
We now use that $R_{12} = R^{-1}_{21} + \lambda {\cal P}$
(and the same for 1$\leftrightarrow$2) to replace one $R$
in each term  in such a way that the terms in ${\cal P}K_{1} R_{21} dK_{2}$
and in ${\cal P}dK_1R_{21}K_2$ may be
cancelled. In this way we obtain two solutions of (\ref{tu}); since the
relations obtained are not  invariant under hermitian conjugation, we may use
one of them
for $dK$ and the other for the hermitian conjugate $dK^{\dagger}$
\begin{equation}\label{tv}
R_{12} K_{1} R_{21} d K_{2} = d K_{2} R_{12} K_{1} R^{-1}_{12} \quad,\quad
R_{12} d K^{\dagger}_{1} R_{21} K_{2} = K_{2} R_{12} dK^{\dagger}_{1}
R_{12}^{-1}  \;,
\end{equation}

\noindent
from which follows that
\begin{equation}\label{tw}
R_{12}dK_{1} R_{21} d K_{2} = -d K_{2} R_{12} dK_{1} R^{-1}_{12} \quad,\quad
R_{12} d K^{\dagger}_{1} R_{21} dK^{\dagger}_{2} = -dK^{\dagger}_{2}
R_{12} dK^{\dagger}_{1}  R_{12}^{-1}  \;.
\end{equation}

\noindent
 We  expect  the $q$-determinant
of $dK$ to vanish; using the first eq. in (\ref{tw})
we check that
\begin{equation}\label{tw2}
tr_q(dK\;dK^{\epsilon}) =0 \;,
\end{equation}
\noindent
where $dK^{\epsilon}$=$\hat{R}^{\epsilon}dK$ (cf. (\ref{dosaf})) and,
in fact, $P_-dK_1 \hat{R} dK_1 P_-=0$.

Finally, to complete the full set of commutation relations, we need those of
$dK$ and $Y$ (and their hermitians). They are given in general by
\begin{equation}\label{tx}
Y_{2} R^{(1)} d K_{1} R^{(2)} = R^{(3)} d K_{1} R^{(4)} Y_{2} \;,
\end{equation}

\noindent
which has the same transformation properties as (\ref{tk}) with $J=0$ and hence
the same solutions (\ref{tm}). Consistency with the previous relations
fixes the solution
\begin{equation}\label{ty}
Y_{2} R^{-1}_{21} dK_1 R_{21} = R_{12} d K_{1} R_{21} Y_{2}\quad, \quad
\tilde{Y}_{2} R_{12} dK^{\dagger}_1 R_{21}
  = R_{12} dK^{\dagger}_{1} R^{-1}_{12} \tilde{Y}_{1} \quad ,
\end{equation}

\noindent
where the second expression corresponds to the only possible consistent
solution for (\ref{tx}) now written for
$\tilde{Y}$ and $dK^{\dagger}$. Notice that eq. (\ref{tx}) is, as any RE,
characterized by the
transformation properties of its entries, and that $Y$ and $\tilde{Y}$ as
well as
$dK$ and $dK^{\dagger}$ transform in the same manner due to the
condition $\tilde{M} = (M^{-1})^{\dagger}$.
With the same structure of (\ref{tx})
(with $Y $ ($dK$) replaced by $\tilde{Y}$ ($dK^{\dagger}$)) and hence with
the same solutions (\ref{tm}),
it is possible to introduce the
commutation properties of $\tilde{Y}, \, dK$ and $Y$, $dK^{\dagger}$.
Consistency gives
\begin{equation}\label{taa}
\tilde{Y}_{2} R^{-1}_{21} dK_1 R_{21} = R_{12} d K_{1} R_{21}
\tilde{Y}_{2}\quad, \quad
Y_{2} R_{12} dK^{\dagger}_1 R_{21}
  = R_{12} dK^{\dagger}_{1} R^{-1}_{12} Y_{1} \quad .
\end{equation}

\noindent
Eqs. (\ref{re}), (\ref{tc}), (\ref{ts}), (\ref{tv}-\ref{tw})
and (\ref{ty}-\ref{taa})  \cite{AKR} define
the full differential calculus  on ${\cal M}_q$ \cite{OSWZ-CMP}.
The identification of the RE algebras generators (entries of $K$, $Y$ and $dK$
matrices) with the ones of ${\cal M}_q$, ${\cal D}_q$ and $\Lambda_q$
of \cite{OSWZ-CMP} is provided by
\begin{equation}\label{matrix}
K\equiv \left[ \begin{array}{cc}
                qD & B \\
                A & C/q
                \end{array} \right] \quad ,\quad
Y\equiv \left[ \begin{array}{cc}
                \partial_D & \partial_A/q \\
                q \partial_B & \partial_C
                \end{array} \right] \quad ,\quad
dK\equiv \left[ \begin{array}{cc}
                qdD & dB \\
               dA & dC/q
                \end{array} \right] \;.
\end{equation}

\vspace{1\baselineskip}

\section{Non-commutative differential calculus and invariant operators}

\indent
 Using the matrices  $K$, $Y$ and $dK$,  we have introduced invariant
differential operators as
the $q$-Minkowski length  $l_{q}$ [(\ref{dosah})], the $q$-D'Alembertian
operator
$\Box_{q}$ [(\ref{ti})] and  the exterior derivative $d$  which,
as its invariance suggests,  has  the form
\begin{equation}\label{tac}
d=  tr_{q} (dKY)\;.
\end{equation}
\noindent
Another one is
a  $q$-analogue of the dilatation operator defined by
\begin{equation}\label{s}
s=tr_q(KY) \quad.
\end{equation}

We conclude by giving below a  list of the expressions
defining the non-commuting properties of
different operators with the corresponding formulae
in  \cite{OSWZ-CMP}. The 4$\times$4 matrices $R$ at the left are those of
$SL_q(2,C)$; the $R$'s at the right are 16$\times$16  matrices.
We refer to \cite{FTUV 94-21}
for more details on the different basis, physical applications,  etc.

\begin{center}
\begin{tabular}{|l||r|}
\hline
\hline
\, \hfill & \hfill \, \\
\, \hspace{1.2cm} {\bf Reflection equations} \hfill  & \hfill
{\bf $q$-Lorentz group$\quad \qquad \quad$} \\
{\bf $\quad SL_q(2,C)$-matrix formulation} \hfill & \hfill
{\bf $R$-matrices formulation
\cite{OSWZ-CMP} $\quad$}
\hspace{0.4cm}\\
\, & \, \\
\hline
\hline
$ R_{12} K_{1} R_{21} K_{2} = K_{2} R_{12} K_{1} R_{21}$ \hfill & \hfill
$ x^{i} x^{j} =  \hat{R}_{I \, kl}^{\; ij} x^{k} x^{l} $ \\
\hline
$R_{12} Y_{1} R_{12}^{-1} Y_{2} = Y_{2} R_{21}^{-1} Y_{1} R_{21} $
\hfill & \hfill
$ \partial_{i} \partial_{j} = \hat{R}_{I \,ji}^{lk} \partial_{k}
\partial_{l} $ \\
\hline
$ Y_{2} R_{12} K_{1} R_{21} = R_{12} K_{1} R_{12}^{-1} Y_{2} + q^{2} R_{12}
{\cal P}$  \hfill & \hfill  $ \partial_{i} x^{j} = \delta_{i}^{j} +
\hat{R}_{II \, il}^{\; jk} x^{l} \partial_{k}$ \\
\hline
$ R_{12} dK_{1} R_{21} dK_{2} = -dK_{2} R_{12} dK_{1} R_{12}^{-1} $
\hfill & \hfill
($\xi^i \equiv dx^i$)  $\;\; \xi^{i} \xi^{j} = - \hat{R}_{II \, kl}^{\; ij}
\xi^{k} \xi^{l}$ \\
\hline
$ R_{12} K_{1} R_{21} dK_{2} = dK_{2} R_{12} K_{1} R_{12}^{-1}$
 \hfill & \hfill
$ x^{i} \xi^{j} = \hat{R}_{II \, kl}^{\; ij} \xi^{k} x^{l}$ \\
\hline
$ Y_{2} R_{21}^{-1} dK_{1} R_{21} = R_{12} dK_{1} R_{21} Y_{2} $
\hfill & \hfill
$\partial_{i} \xi^{j} = \hat{R}_{II\, il}^{-1\, jk} \xi^{l}
\partial_{k} $ \\
\hline
$ R_{21}^{-1} Y_{1} R_{12}^{-1} \tilde{Y}_{2} = \tilde{Y}_{2} R_{21}^{-1} Y_{1}
R_{21} $ \hfill & \hfill $ \partial_{i} \hat{\partial}_{j} =
\hat{R}_{II\, ji}^{-1\, lk}
\hat{\partial}_{k} \partial_{l}$ \\
\hline
$ \tilde{Y}_{2} R_{21}^{-1} K_{1} R_{21} = R_{12} K_{1} R_{21} \tilde{Y}_{2} +
q^{-2} R_{12} {\cal P} $ \hfill & \hfill $ \hat{\partial}_{i} x^{j} =
\delta_{i}^{j}
+ \hat{R}_{II\, il}^{-1\, jk} x^{l} \hat{\partial}_{k}$ \\
\hline
$ R_{21}^{-1} K_{1} R_{21} dK^{\dagger}_{2} = dK^{\dagger}_{2} R_{12} K_{1}
R_{21}
$ \hfill & \hfill
$ x^{i} \hat{\xi}^{j} = \hat{R}_{II\, kl}^{-1\, ij} \hat{\xi}^{k} x^{l}$\\
\hline
\hline
\,$ l_{q} \equiv  \frac{-q}{[2]} tr_{q} (KzK^{\epsilon}) $ \hfill & \hfill $
(q^{2} + 1)^{-1} g_{ij} x^{i}x^{j} = AB - q^{-2} CD \equiv L $ \\
\hline
$ \Box_{q} = \frac{-q^{-1}}{[2]} tr_{q} (YzY^{\epsilon}) $ \hfill & \hfill
$(q^{2} + 1)^{-1}
g^{ij} \partial_{j} \partial_{i} = \partial_{A} \partial_{B} - q^{2}
\partial_{C} \partial_{D} \equiv \Delta$ \\
\hline
$ 0 = tr_{q} (dK z dK^{\epsilon}) $ \hfill & \hfill
$g_{ij} \xi^{i} \xi^{j} = 0 $ \\
\hline
$Y l_{q} = q^{-2} l_{q} Y - q^{2} K^{\epsilon} $ \hfill & \hfill $
\partial_{i} L = q^{-2} L \partial_{i} + g_{ij} x^{j} $ \\
\hline
$l_{q} z dK = q^{-2} dK z l_{q} $ \hfill & \hfill $ L \xi^{i} = q^{-2}
\xi^{i} L $ \\
\hline
$l_{q} \Box_{q} = q^{-4} l_{q} \Box_{q} + q^{-2}s + (q^{2}+1) $
\hfill &
\hfill $ \Delta L = q^{-4} L \Delta + q^{-2} E + (q^{2}+1)$ \\
\hline
$ \Box_{q} K = q^{-2} K \Box_{q} - Y^{\epsilon} $ \hfill & \hfill
$\Delta x^{i} = q^{-2}
x^{i} \Delta + q^{2} g^{ij} \partial_{j}$\\
\hline
$\Box_{q} dK = q^{2} dK \Box_{q} $ \hfill & \hfill $
\Delta \xi^{i} = q^{2} \xi^{i} \Delta $ \\
\hline
\hline
$d = tr_{q} (dK \, Y) $ \hfill & \hfill  $d = \xi^{i} \partial_{i}$\\
\hline
$ d \cdot K = (dK) + K z d $ \hfill & \hfill $
d z x^{i} = \xi^{i} + x^{i} d$ \\
\hline
$Y z d = q^{2} d z Y + (q^2-1)  dK^{\epsilon} \Box_{q}$ \hfill  & \hfill $
\partial_{i} d =
q^{2} d \partial_{i} -(1-q^{-2}) g_{ij} \xi^{j} \Delta$ \\
\hline
$ d \, (dK) = -(dK) \, d $ \hfill & \hfill $ d z \xi^{i} = -\xi^{i} d$ \\
\hline
$ d z \Box_{q} = q^{-2} \Box_{q} z d $ \hfill & \hfill $ d z \Delta =
q^{-2} \Delta d$ \\
\hline
$ d \, l_{q} = l_{q} d-q^2W \quad(W \equiv tr_q (dK z K^{\epsilon}))$
\hfill & \hfill $ d z L = L d+W \quad (W \equiv  g_{ij} \xi^{i} x^{j}) $ \\
\hline
\hline
$ s \equiv tr_{q} (KY) $ \hfill  & \hfill
$ E \equiv x^{i} \partial_{i}$ \\
\hline
$ s K = q^{-2} K s + K - q^{-1} \lambda l_{q} Y^{\epsilon} $ \hfill & \hfill
$E x^{i} = q^{-2} x^{i} E + x^{i} + q \lambda L g^{ij} \partial_{j}$\\
\hline
$Y s= q^{-2} s Y + Y - q \lambda K^{\epsilon} \Box_{q} $ \hfill & \hfill
$ \partial_{i} E = q^{-2} E \partial_{i} + \partial_{i} + q^{-1} \lambda g_{ij}
x^{j} \Delta$ \\
\hline
$s\,dK=dK\,s$ \hfill & \hfill $E\, \xi^i = \xi^i \, E$ \\
\hline
$ sl_{q} = q^{-2} l_{q} s + (q^{2} + 1) l_{q} $ \hfill & \hfill
$EL= q^{-2} LE + (q^{2}+1)L $\\
\hline
$d\,s=d+ q^{-2}s\,d  -q \lambda W \Box_q $ \hfill & \hfill
$d\,E= d+ q^{-2}E\,d+ q^{-1} \lambda W \Delta$ \\
\hline
\hline
\end{tabular}
\end{center}

\vspace{1\baselineskip}

\noindent
{\bf Acknowledgements}: The authors wish to thank P.P. Kulish for helpful
discussions.
This research has been  partially supported by
a CICYT research grant.

\end{document}